\documentclass[journal=ancac3,manuscript=article]{achemso}

\usepackage{graphicx}
\usepackage[pdfencoding=auto, psdextra]{hyperref}
\usepackage{amsmath}
\usepackage{amssymb}
\usepackage[normalem]{ulem}
\usepackage[usenames,dvipsnames]{color}
\usepackage{float}
\usepackage{xcolor}
\usepackage{soul}

\graphicspath{{figures/}}

\author{Linghao Yan}
\email{linghao.yan@aalto.fi}
\affiliation{Department of Applied Physics, Aalto University, 00076 Aalto, Finland}
\altaffiliation{Contributed equally to this work}

\author{Orlando J. Silveira}
\affiliation{Department of Applied Physics, Aalto University, 00076 Aalto, Finland}
\altaffiliation{Contributed equally to this work}

\author{Benjamin Alldritt}
\affiliation{Department of Applied Physics, Aalto University, 00076 Aalto, Finland}

\author{Shawulienu Kezilebieke}
\affiliation{Department of Applied Physics, Aalto University, 00076 Aalto, Finland}

\author{Adam S. Foster}
\affiliation{Department of Applied Physics, Aalto University, 00076 Aalto, Finland}
\alsoaffiliation{Nano Life Science Institute (WPI-NanoLSI), Kanazawa University, Kakuma-machi, Kanazawa 920-1192, Japan}

\author{Peter Liljeroth}
\email{peter.liljeroth@aalto.fi}
\affiliation{Department of Applied Physics, Aalto University, 00076 Aalto, Finland}

\title{A Two-Dimensional Metal-Organic Framework on Superconducting \texorpdfstring{NbSe$_2$}{NbSe2}}

\keywords{electronic structures, metal-organic framework, scanning tunneling microscopy, tunneling spectroscopy, on-surface synthesis, 2D material}

\begin{document}

\begin{abstract}

The combination of two-dimensional (2D) materials into vertical heterostructures has emerged as a promising path to designer quantum materials with exotic properties. Here, we extend this concept from inorganic 2D materials to 2D metal-organic frameworks (MOFs) that offer additional flexibility in realizing designer heterostructures. We successfully fabricate a monolayer 2D Cu-dicyanoanthracene MOF on a 2D van der Waals NbSe$_2$ superconducting substrate. The structural and electronic properties of two different phases of the 2D MOF are characterized by low-temperature scanning tunneling microscopy (STM) and spectroscopy (STS), complemented by density-functional theory (DFT) calculations. These experiments allow us to follow the formation of the kagome bandstructure from Star of David -shaped building blocks. This work extends the synthesis and electronic tuneability of 2D MOFs beyond the electronically less relevant metal and semiconducting surfaces to superconducting substrates, which are needed for the development of emerging quantum materials such as topological superconductors.

\end{abstract}


\section{Introduction}

Two-dimensional (2D) materials have attracted broad attention because of their outstanding properties and wide range of material properties that can be realized \cite{Liu2016,Du2021}. The properties of the individual materials can be further developed in van der Waals (vdW) heterostructures, exploiting the interactions between layers to fabricate designer systems based on exotic electronic properties. This is exemplified by twisted bilayer graphene samples exhibiting superconductivity and correlated insulator states \cite{Andrei2021,Kennes2021}. There are also examples of using vdW materials to realize exciton condensates, quantum spin liquids, Chern insulators and topological superconductivity in vdW heterostructures \cite{Wang2019,PhysRevLett.124.106804,Nuckolls2020,Liu2021,Andrei2021,Kezilebieke2020}.

While recent research in realizing vdW heterostructures has focused on inorganic 2D materials, 2D metal-organic frameworks (MOF) form an extremely interesting, broad, and tunable class of materials. MOFs are well-established in topics such as single-atom catalysis or gas storage, but there is also growing interest in the intrinsic electronic properties of 2D MOFs. Theoretical works have predicted their usage to realize, for example, 2D topological insulators \cite{Silveira2017,Springer2020,Jiang2021}, half-metallic ferromagnetism \cite{Zhao2013Half-metallicityMonolayer,Zhang2015RobustCu-TPyB,Jin2018Large-gapFrameworks,Zhang2019Two-dimensionalLattice} and quantum spin liquids \cite{Yamada2017DesigningFrameworks,Misumi2020}. 

2D MOFs have been synthesized on metal surfaces by following the concepts of supramolecular coordination chemistry \cite{Lin2009,Barth2009FreshChemistry}. So far, most of the 2D MOFs are made on coinage metal surfaces \cite{Dong2016Self-assemblySurfaces,Gao2019SynthesisGap,Zhang2020OnsurfaceBands,Gao2020,Hua2021}, where the interaction with the metal substrate strongly masks the intrinsic electronic properties of the MOF. This can be overcome by using weakly interacting substrates (such as graphene and hBN) that allow probing of the intrinsic exotic electronic properties of 2D MOFs \cite{Urgel2015ControllingMonolayer,Kumar2018Two-DimensionalFrameworks,Zhao2019OnSurfaceTemplates,Li2019,Yan2021,Zhang2016IntrinsicLattices}. However, to realize the exciting prospect of truly designer materials, it is important to demonstrate MOF synthesis on other 2D substrates.

Among possible candidates of MOF-related designer materials, MOFs on superconductors are particularly interesting. Magnetic adsorbates on a superconducting surface give rise to the Yu-Shiba-Rusinov (YSR) states \cite{Yu1965,Shiba1968,Rusinov1969,Heinrich2018}. Furthermore, 2D magnetic lattices on a superconductor (YSR lattice) can lead to intriguing 2D topological superconductivity \cite{Rontynen2015,Li2016,Rachel2017,Menard2017,Palacio-Morales2019,Kezilebieke2020}. The YSR states have been observed on individual 3$d$ transition metal-phthalocyanine molecules on superconducting substrates, including layered vdW material NbSe$_2$ \cite{Franke2011,Kezilebieke2018,Heinrich2018,Kezilebieke2019,Liebhaber2020}. While MOFs have not been realized on superconducting transition metal dichalcogenides, Ahmadi \emph{et al.}~have fabricated Pb-TNAP and Na-TNAP networks on a Pb surface \cite{Ahmadi2016}. However, the insufficient mobility of the adsorbates hinders the formation of an ordered transition metal-based MOF \cite{Ahmadi2016}.

In this work, we successfully fabricated 2D Cu-dicyanoanthracene (DCA) MOF on a NbSe$_2$ superconducting substrate under ultra-high vacuum (UHV) conditions. The structural and electronic properties of the samples are studied by low-temperature scanning tunneling microscopy (STM) and spectroscopy (STS). The ordered DCA$_3$Cu$_2$ network shows a structure that is a combination of a honeycomb lattice of Cu atoms with a kagome lattice of DCA molecules. Interestingly, we observed an unexpected Star of David (SD) lattice phase after further annealing the sample at room temperature. The evolution of energy bands from one SD unit cell to the kagome band of 2D MOF was observed by comparing the STS of these two phases combined with density-functional theory (DFT) calculations. Given that a similar magnetic transition metal-based 2D MOF has been successfully fabricated on a weakly interacting graphene/Ir(11) substrate, we think that the same strategy can be easily applied to the synthesize of 2D MOFs with magnetic transition atoms such as Co \cite{Kumar2018Two-DimensionalFrameworks} on a van der Waals superconductor -- possibly leading to topological superconductivity.

\section{Results and Discussion}

\begin{figure}[ht]
    \includegraphics[width=0.95\textwidth]{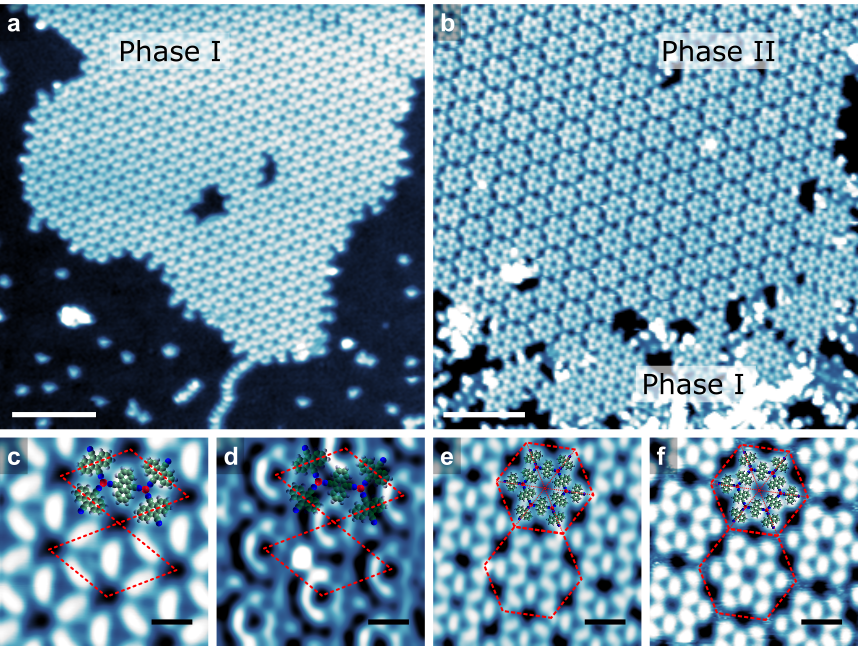}\\
    \centering
    \caption{(a,b) STM overview images of Cu-DCA MOFs on the NbSe$_2$ surface. (a), phase~\uppercase\expandafter{\romannumeral1}, (b), phases~\uppercase\expandafter{\romannumeral1} and~\uppercase\expandafter{\romannumeral2}. (c,d) STM images of phase~\uppercase\expandafter{\romannumeral1}. (e,f) STM images of phase II. The red parallelograms indicate the unit cells (C: cyan, N: blue, H: white, Cu: red). Imaging parameters: (a) 1.0 V and 2 pA, (b) 1.0 V and 10 pA, (c) 1.0 V and 11 pA, (d) 1.8 V and 11 pA, (e) 0.1 V and 10 pA, (f) 1.0 V and 10 pA. Scale bars: (a) and (b) 10 nm, (c) and (d) 1 nm, (e) and (f) 2 nm.}
    \label{stm}
\end{figure}

We deposited DCA molecules and Cu atoms sequentially onto the NbSe$_2$ substrate held at room temperature (details given in the Methods Section). In a stark contrast with the close-packed assembly of DCA molecules shown in Figure~S1, an ordered DCA$_3$Cu$_2$ network (denoted as phase~\uppercase\expandafter{\romannumeral1}) was observed in Figure~\ref{stm}a, where the Cu atoms arrange in a honeycomb lattice and the DCA molecules form a kagome pattern. Figures~\ref{stm}c and \ref{stm}d show high-resolution images of DCA$_3$Cu$_2$ network. Figure~\ref{stm}c shows the backbone of the network, with the unit cell shown as a red parallelogram with a lattice constant of $a=1.96$ nm, which is in the range of previous reports of the same network on other substrates \cite{Pawin2008AExcess,Zhang2014ProbingNetwork,Hernandez-Lopez2021,Yan2021,KumarAgustin2021} and comparable with the gas phase DFT values (2.02 nm). The electronic states of the DCA$_3$Cu$_2$ can be seen at a higher bias (1.8 V) shown in Figure~\ref{stm}d (see below for more detailed spectroscopy of the network electronic structure).

After further annealing the sample at room temperature, phase~\uppercase\expandafter{\romannumeral2} emerged as the dominant pattern (see Figure~\ref{stm}b), where the DCA molecules formed a SD lattice, indicating that phase~\uppercase\expandafter{\romannumeral1} is metastable.  Figures~\ref{stm}e and \ref{stm}f are the high-resolution images of phase~\uppercase\expandafter{\romannumeral2}. At 0.1 V (Figure~\ref{stm}e), it can be seen that there are bright dots between each SD in phase~\uppercase\expandafter{\romannumeral2}, while at 1.0 V (Figure~\ref{stm}f) those areas become fuzzy. The SD unit cell shown as a red hexagon in Figure~\ref{stm}e has a lattice constant of $a=4.33$ nm, which is 10\% larger than double the lattice constant of phase~\uppercase\expandafter{\romannumeral1}.
Note that phase~\mbox{\uppercase\expandafter{\romannumeral1}} undergoes a 5\% lattice mismatch with the lattice constant of a 6$\times$6 NbSe$_2$, while the distance of the centers of outer DCA molecules in a single cell of phase~\mbox{\uppercase\expandafter{\romannumeral2}} (3.46 nm) almost perfectly matches the lattice constant of a 10$\times$10 NbSe$_2$ (0.5\% mismatch) thus is more favorable after annealing. This appears to be coverage independent, as the experiments are always at sub-monolayer coverage.

\begin{figure}[ht]
    \includegraphics[width=0.95\textwidth]{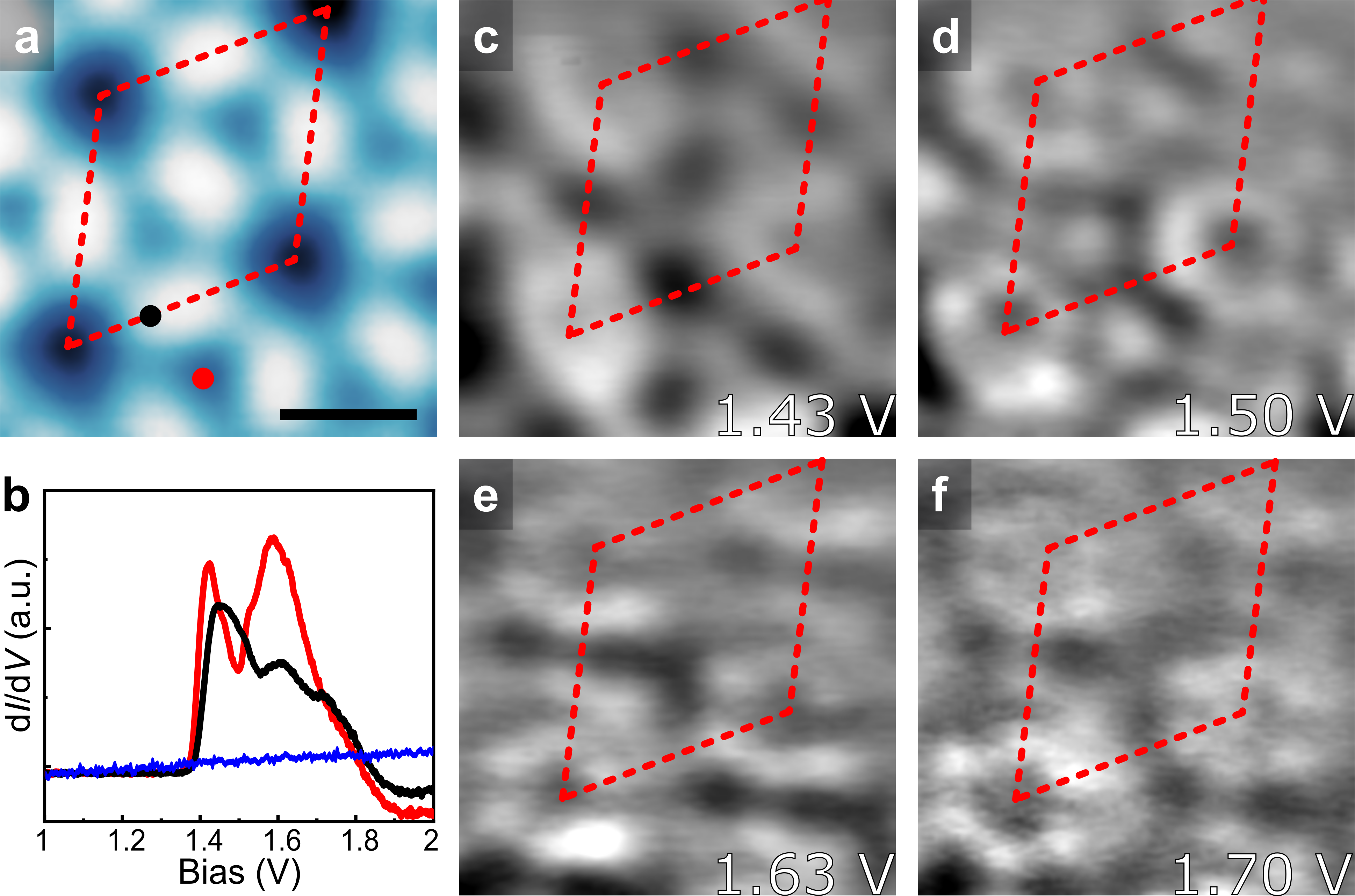}\\
    \centering
    \caption{(a,b) STS (b) on phase~\uppercase\expandafter{\romannumeral1} at the positions shown in (a), the blue curve in (b) is a reference spectrum on bare NbSe$_2$. (c-f) Experimentally recorded constant-height d$I$/d$V$ maps at the energies indicated in the panels in the same area of (a). Imaging parameters of (a): 1.0 V and 10 pA. Scale bar of (a): 1 nm.}
    \label{kagome}
\end{figure}

The d$I$/d$V$ spectra recorded on a DCA molecule and Cu atom in phase~\uppercase\expandafter{\romannumeral1} are shown in Figure~\ref{kagome} and Figure~S2. It can be seen that there is a region of negative differential conductance directly above the band of MOF. This is not an intrinsic electronic feature of the 2D MOF but arises from the bias dependence of the tunneling barriers.\cite{Repp2005_1,Grobis2005,Repp2005_2} Both spectra exhibit a broad feature in the energy range between 1.4 V to 1.9 V. We recorded constant height d$I$/d$V$ maps at representative biases indicated in Figures~\ref{kagome}c-f, which show nearly uniform features at different biases. These characteristics are attributed to the band structure formed in the 2D MOF, which has been well studied in the same network on graphene/Ir(111) \cite{Yan2021}. It is worth noting that the energy position of this band shifts from close to the Fermi level on graphene/Ir(111) to 1.4 V on NbSe$_2$. The energy band shift can be primarily explained by the different work functions (\emph{i.e.,} the energy difference between the vacuum level and the Fermi energy) of graphene/Ir(111) ($4.65 \pm 0.10$ eV) \cite{Niesner2012} and NbSe$_2$ (5.9 eV) \cite{Shimada1994}. Due to the instability of tunneling junction at negative bias, we were only able to get reliable STS results at biases above -0.5~V. We carried out DFT calculations considering phase~\uppercase\expandafter{\romannumeral1} DCA$_3$Cu$_2$ on monolayer NbSe$_2$, where the stacking geometries can be seen in Figure~S3. Figure~S4 shows that the band structure of the MOF in this heterostructure is 0.6 eV higher relative to the Fermi level, in qualitative agreement with our experimental findings.

\begin{figure}[ht]
    \includegraphics[width=0.95\textwidth]{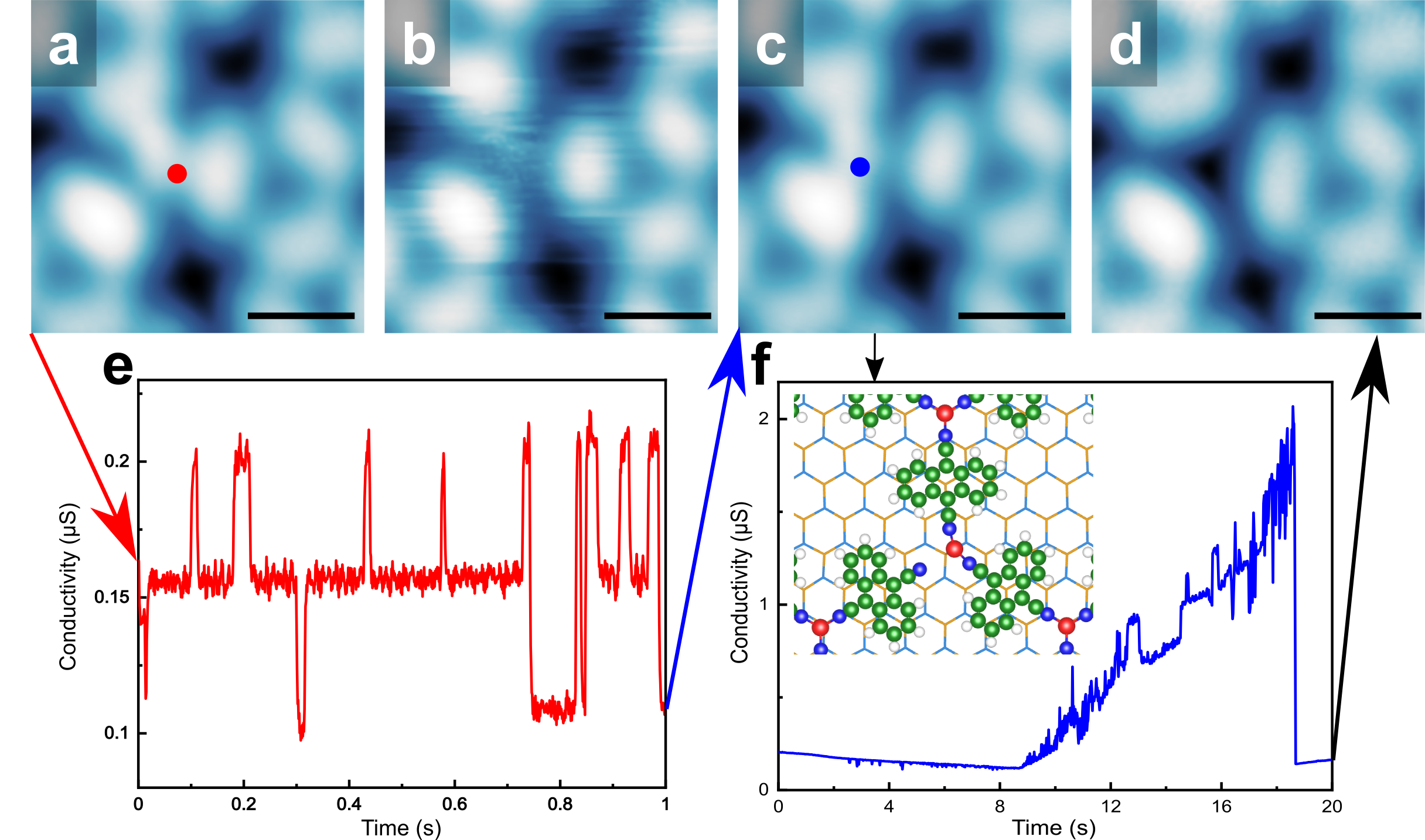}\\
    \centering
    \caption{(a-d) STM images of phase~\uppercase\expandafter{\romannumeral2} recorded before and after measuring the current \emph{versus}  time spectra. (e) Current \emph{versus}  time trace recorded at 0.4 V at the position indicated by the red dot. (f) Current \emph{versus}  time trace recorded at 3.0 V at the position indicated by the blue dot in c. Imaging parameters: (a) 0.1 V and 10 pA, (b) 0.8 V and 10 pA, (c) 0.1 V and 10 pA, (d) 0.1 V and 10 pA. Scale bars: (a-d) 1 nm. The inset in (f) shows the outer Cu atom binding to two of the three N atoms of the correspondent DCA molecules (see also Figure~S5), where the wireframe in the background represents the NbSe$_2$ substrate.}
    \label{dstarit}
\end{figure}

The Cu-N bond is estimated to be $2.0 \pm 0.2$ \AA~within the SD, in line with our DFT result ($1.9$ \AA) and with the value of the earlier report of Cu-N coordination bond \cite{Su1994}. In contrast, the distance between the fuzzy center and the surrounding N atom is $3.5 \pm 0.3$ \AA, which beyond the range of a typical Cu-N bond. It can be seen from the images at low bias (0.1V, Figures~\ref{dstarit}a and c) that the bright dot was not located at the center of the three surrounding molecules, but only attached to two of the molecules. While at the bias of 0.8 V, the bright dot area became unstable and thus results in a fuzzy feature in Figure~\ref{dstarit}b, which also affects the surrounding DCA molecules. We measured the current \emph{versus}  time ($I-t$) traces at the positions indicated in Figures~\ref{dstarit}a and c. While using a relatively low bias (0.4 V) for the $I-t$ spectrum, a switching between three current levels was clearly observed, as shown in Figure~\ref{dstarit}e. A plausible explanation would be the Cu atom hopping between the three possible binding geometries. When using a relative high bias (3.0 V), the tunneling current became unstable and increased dramatically until a sharp drop occurred. The sharp drop in current was likely because the Cu atom was picked up by the tip, which was further evidenced by the STM image taken after acquiring the $I-t$ trace (Figure~\ref{dstarit}d). The inset of Figure~\ref{dstarit}f shows a portion of the phase~\uppercase\expandafter{\romannumeral2} DCA$_3$Cu$_2$ on a monolayer NbSe$_2$ obtained through our DFT calculations, which shows that the outer Cu atom of the SD will only bond to two N atoms in this scenario (see also Figure~S5), which agrees with our experimental findings. In this case, the shortest distance measured between the Cu atom and the three surrounding N atoms is $1.9$ \AA, agreeing well with the Cu-N typical bond length, while the largest distance measured is $3.7$ \AA, again matching experiments.

\begin{figure}[ht]
    \includegraphics[width=0.95\textwidth]{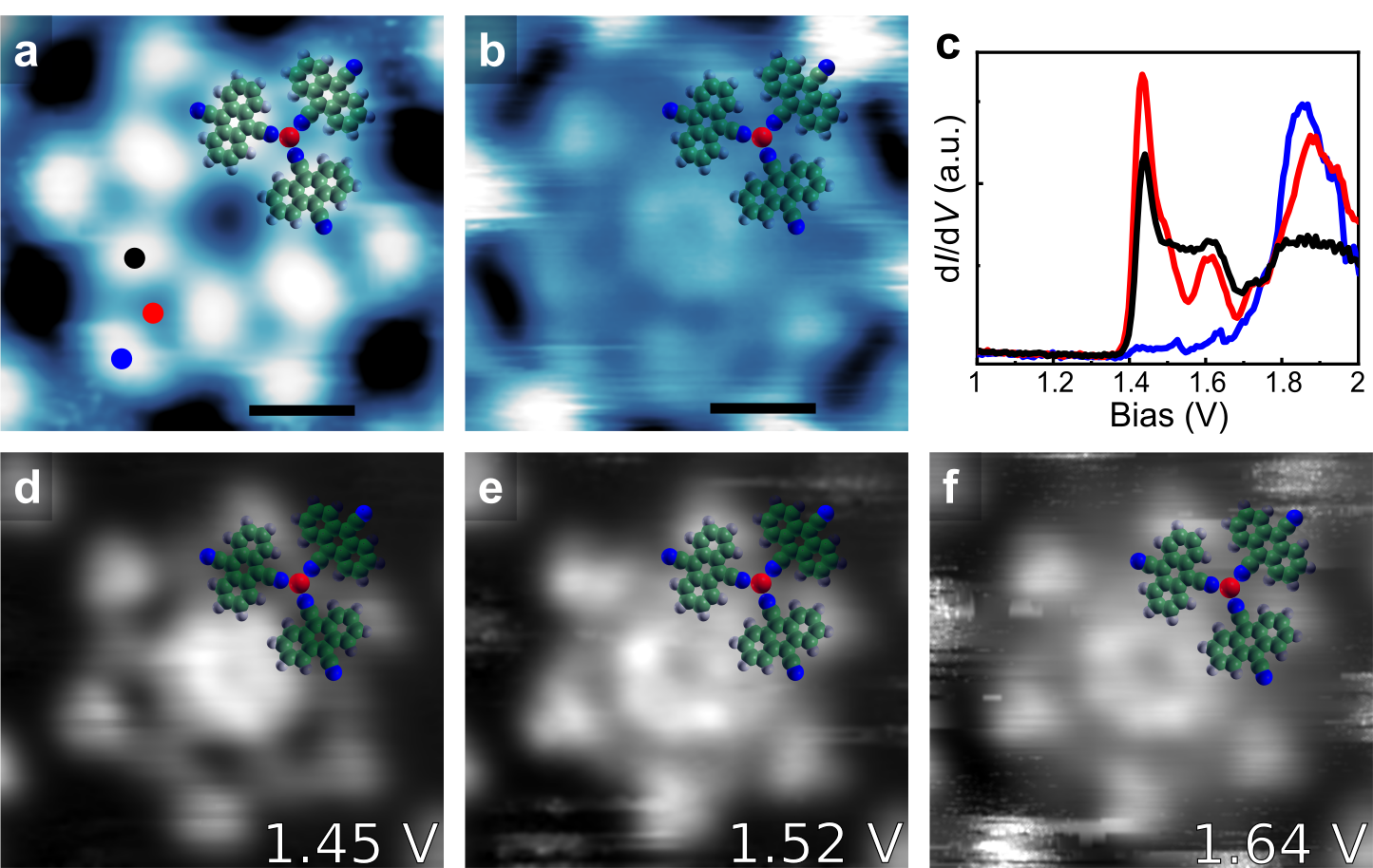}\\
    \centering
    \caption{(a-c) STS recorded (c) on phase~\uppercase\expandafter{\romannumeral2} at the positions shown in (a). (d-f) Experimentally recorded constant-height d$I$/d$V$ maps at the energies indicated in the panels in the same area of (a) and (b). Imaging parameters: (a) 1.0 V and 10 pA, (b) 2.0 V and 10 pA. Scale bars: (a,b) 1 nm.}
    \label{dstar}
\end{figure}

The electronic properties of phase~\uppercase\expandafter{\romannumeral2} are shown in Figure~\ref{dstar} and Figure~S6. While the d$I$/d$V$ spectra recorded on the Cu atom and the inner DCA molecule still show the same broad feature similar to phase~\uppercase\expandafter{\romannumeral1} with two peaks at 1.45 V and 1.64 V, the outer DCA molecule has very limited local density of states (LDOS) at these energies, but shows a strong peak at 1.85 V, which is reflected on the neighboring Cu atom (the nearby CN group of the outer DCA molecule) as well. The constant height d$I$/d$V$ maps below 1.6 V (Figures~\ref{dstar}d and e) show LDOS intensity only on the inner DCA molecules and Cu atoms. At 1.64 V, while the character of the inner parts remain the same in the constant height d$I$/d$V$ map (Figures~\ref{dstar}f), the outer part becomes fuzzy due to Cu atom instability, as discussed above. An STM image recorded at 2.0 V (Figures~\ref{dstar}b) shows the electronic contributions consisting of all the 
states from 0 V to 2 V of phase~\uppercase\expandafter{\romannumeral2}, where contrast within the SD is dominated by the ends of the long-axis of the DCA molecules. 

\begin{figure}[ht]
    \includegraphics[width=0.95\textwidth]{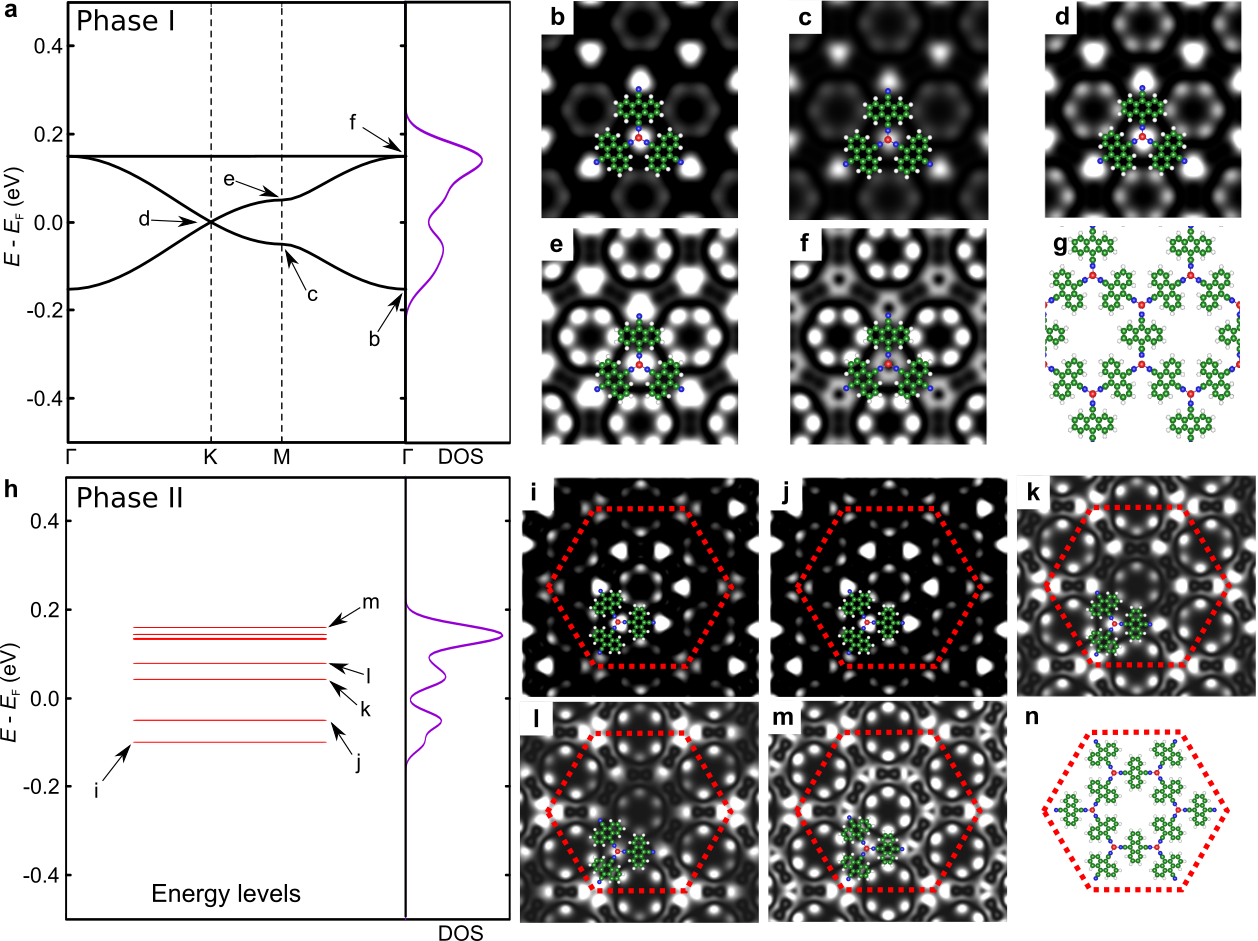}\\
    \centering
    \caption{Calculated band structure, energy levels, DOS, simulated LDOS maps at the energies indicated in the band structure, and the related model of gas-phase Cu-DCA MOF. Upper panels: phase~\uppercase\expandafter{\romannumeral1}. Lower panels: phase~\uppercase\expandafter{\romannumeral2}. The red dashed lines in the lower panels indicate the SD unit cell of phase~\uppercase\expandafter{\romannumeral2}.
    }
    \label{DFT}
\end{figure}

The differences in the electronic properties between phases~\uppercase\expandafter{\romannumeral1} and~\uppercase\expandafter{\romannumeral2} can also be observed in their gas-phase state, as shown in our DFT simulations. Figure~5a shows the calculated band structure of the pristine, ordered DCA$_3$Cu$_2$ network (phase~\uppercase\expandafter{\romannumeral1}), revealing a kagome band structure which consists of a Dirac band with an additional flat band pinned to the top of the Dirac band \cite{Leykam2018a, Yan2019EngineeredNanoribbons,Jing2019Two-DimensionalSemiconductors}. We simulated the constant-height d$I$/d$V$ maps by extracting the local density of states (LDOS) maps on selected points of the kagome band structure, as can be seen in Figures~\ref{DFT}b-g. Overall, the simulations are consistent with our experimental findings. The LDOS maps in Figures~\ref{DFT}b-d show that higher contrast is observed on the N atoms and the end of the axis of the DCA molecules, with a minor contribution from the Cu atoms (see also Figure~S4), where the LDOS maps were extracted from energies within the Dirac band-like spectrum of the kagome band. At higher energies closer to the flat band (Figures~\ref{DFT} e-f), the LDOS contrast on the DCA molecules is significantly increased, which is consistent with the flat band arising from the entities located at the kagome sites of the unit cell \cite{Yan2021}. Note again that the kagome band is observed around the Fermi level in the gas-phase calculation, whereas in the experiment these features are observed 1.4 eV above the Fermi level. We stress here that calculations considering the NbSe$_2$ substrate also gives these features well above the Fermi level (0.6 eV), which agrees qualitatively with the experiment. However, the fact that the gas-phase calculation accurately captures the electronic properties of the DCA$_3$Cu$_2$ MOF, shows that the layers (MOF and NbSe$_2$) do not interact strongly with each other and retain their intrinsic properties in the heterostructure. 

For phase~\uppercase\expandafter{\romannumeral2}, as a simple approximation to simulate its electronic properties, we considered a periodic SD lattice fixing the experimental lattice parameter (4.33 nm) without the outer Cu atoms. Therefore, the distance between the SD units is large enough to isolate them, resulting in molecular-like energy levels rather than a dispersive band structure, such as for phase~\uppercase\expandafter{\romannumeral1}. Nonetheless, the DOS of phase~\uppercase\expandafter{\romannumeral2} shows similar characteristics to phase~\uppercase\expandafter{\romannumeral1} since the lattice hosts a complete SD unit cell. The simulated LDOS maps of phase~\uppercase\expandafter{\romannumeral2} at low bias (Figures~\ref{DFT}i and j) show higher contrast on the inner entities, while at high bias (Figure~\ref{DFT}k-m) the outer DCA molecules (especially the outer N atoms) possess stronger LDOS. These characteristics are in agreement with the experiment, where the outer DCA molecules have stronger DOS at higher bias (Figures~\ref{dstar}c). In other words, the comparison between the two different phases allows us to visualize the evolution of energy bands from one SD unit cell to the kagome band of a 2D MOF.

\section{Conclusions}
In summary, we have studied the structural and electronic properties of two different phases of 2D Cu-DCA MOF on a NbSe$_2$ substrate under UHV conditions using experimental (STM/STS) and theoretical (DFT) methods. Phase~\uppercase\expandafter{\romannumeral1} is an ordered DCA$_3$Cu$_2$ network possessing a kagome band structure. Phase~\uppercase\expandafter{\romannumeral2} consists of a lattice of an electronically isolated Star of David (SD) network with unstable Cu atoms between them. The differences in the electronic structures of the two phases can be understood in terms of an evolution of the energy bands from one SD unit cell to the kagome band of 2D MOF. This work demonstrates the successful synthesis of a 2D MOF on the superconducting NbSe$_2$ substrate capable of producing 2D MOF-superconductor hybrids with the proposed magnetic \cite{Kumar2018Two-DimensionalFrameworks} or organic topological insulator MOFs \cite{Zhang2016IntrinsicLattices,Lyu2015On-surfaceStructures,Yan2018StabilizingNetworks,Yan2019On-SurfaceAu111}.

\section{Methods}
\subsection{Experimental}
Sample preparation and STM experiments were carried out in an ultrahigh vacuum system with a base pressure of $\sim 10^{-10}$ mbar. The 2H–NbSe2 single crystal (HQ Graphene, the Netherlands) was cleaved \emph{in-situ} in the vacuum \cite{Kezilebieke2018}. The DCA$_3$Cu$_2$ network was fabricated by the sequential deposition of 9,10-dicyanoanthracene (DCA, Sigma Aldrich) molecules (evaporation temperature 100 $^\circ$C) and Cu atoms onto the NbSe$_2$ substrate held at room temperature. The star phase emerged after annealing 10 hours at room temperature or 5 minutes at 40 $^\circ$C. Subsequently, the samples were inserted into the low-temperature STM (Createc GmbH), and all subsequent experiments were performed at $T = 5$ K. STM images were recorded in constant current mode. d$I$/d$V$ spectra were recorded by standard lock-in detection while sweeping the sample bias in an open feedback loop configuration, with a peak-to-peak bias modulation of 15 - 20 mV at a frequency of 526 Hz. The STM images were processed with Gwyddion software \cite{Necas2012Gwyddion:Analysis}.

\subsection{Computational}
DFT calculations were performed using the FHI-AIMS code \cite{BLUM20092175}. The default calculation setup used a ``tight" basis set and the Perdew-Burke-Ernzerhof (PBE) exchange-correlation functional \cite{PhysRevLett.77.3865} augmented with van der Waals terms through the Tkatchenko-Scheffler method \cite{PhysRevLett.102.073005}. Within this methodology, different strategies were applied for the gas phase and on surface calculations, which are described as follows.

\noindent \textbf{Gas-phase:} For phase~\uppercase\expandafter{\romannumeral1} DCA$_3$Cu$_2$, the geometry was fully relaxed considering a 2$\times$2$\times$1 uniform $k$-grid, while for phase~\uppercase\expandafter{\romannumeral2} we kept the experimental lattice parameter fixed and a single $\Gamma$ $k$-point was used. Electronic properties such as band structures, energy levels, density of states, as well as the LDOS maps were obtained by partially including exact exchange terms into the exchange-correlation functional by means of the hybrid Heyd-Scuseria-Ernzerhof (HSE06) functional \cite{Ren_2012,doi:10.1063/1.1564060,doi:10.1063/1.2204597}, which has been shown to be relevant for describing the electronic properties of MOFs \cite{Silveira2016,Silveira2017}. A significantly larger $k$-grid was used for the band structures and density of states calculations. LDOS maps were computed by means of the PP-STM code \cite{PhysRevB.95.045407}, where the broadening parameter $\eta$ was set to 0.2 eV. 

\noindent \textbf{On surface:} For the phase~\uppercase\expandafter{\romannumeral1} DCA$_3$Cu$_2$ on a NbSe$_2$, the DCA$_3$Cu$_2$ was freely relaxed on a fixed 6$\times$6 supercell NbSe$_2$ monolayer with a 2$\times$2$\times$1 uniform k-grid. A strain of 3\% was present on phase~\uppercase\expandafter{\romannumeral1} DCA$_3$Cu$_2$ due to the mismatch between its lattice parameter and NbSe$_2$. The hybrid HSE06 functional was used to obtain the electronic properties. For phase~\uppercase\expandafter{\romannumeral2} DCA$_3$Cu$_2$ on NbSe$_2$, the DCA$_3$Cu$_2$ was freely relaxed on a fixed 12$\times$12 supercell NbSe$_2$ monolayer.

In this case we used the "light" basis set due to the large size of the system. 
Figures~S3 and S5 show the unit cell, atomic structures and stacking configuration of all structures considered here.

\begin{acknowledgement}
This research made use of the Aalto Nanomicroscopy Center (Aalto NMC) facilities. We acknowledge support from the European Research Council (ERC-2017-AdG no.~788185 ``Artificial Designer Materials'') and Academy of Finland (Academy projects no.~311012 and 314882, and Academy professor funding no.~318995 and 320555), and Academy Research Fellow no.~338478 and 346654. L.Y. acknowledges support from European Union's Horizon 2020 research and innovation program (Marie Skłodowska-Curie Actions Individual Fellowship no.~839242 ``EMOF''). Computing resources from the Aalto Science-IT project and CSC, Helsinki are gratefully acknowledged. A.S.F has been supported by the World Premier International Research Center Initiative (WPI), MEXT, Japan.
\end{acknowledgement}

\bibliography{Ref}

\end{document}